\documentclass[%
 reprint,
 amsmath,amssymb,
 aps,
 prl,
]{revtex4-1}

\usepackage{graphicx}
\usepackage{dcolumn}
\usepackage{bm}

\usepackage{color}

\newcommand{\down}{\downarrow}
\newcommand{\up}{\uparrow}

\begin{document}

\title{Raising the critical temperature by disorder in unconventional superconductors mediated by spin fluctuations}

\author{Astrid T. R\o mer,$^1$ P. J. Hirschfeld,$^2$ Brian M. Andersen$^1$}
\affiliation{%
$^1$Niels Bohr Institute, University of Copenhagen, Juliane Maries Vej 30, DK-2100 Copenhagen,
Denmark\\
$^2$Department of Physics, University of Florida, Gainesville, Florida 32611, USA
}%

\date{\today}

\begin{abstract}
{ 
We propose a mechanism whereby disorder can enhance the transition temperature $T_c$ of an unconventional superconductor with pairing driven by exchange of spin fluctuations. The theory is based on a self-consistent real space treatment of pairing in the disordered one-band Hubbard model. It has been demonstrated before that impurities can enhance pairing by softening the spin fluctuations locally; here, we consider the competing effect of pair-breaking by the screened Coulomb potential also present. We show that, depending on the impurity potential strength and proximity to magnetic order, this mechanism  results in a weakening of the disorder-dependent $T_c$-suppression rate expected from Abrikosov-Gor'kov theory, or even in disorder-generated $T_c$ enhancements. }

\end{abstract}

\maketitle


{\it Introduction.}  Disorder has been used as a powerful probe of superconducting order since a theoretical framework for interpreting its effects was provided by Anderson\cite{Anderson59} and Abrikosov-Gor'kov\cite{ag} (AG).  Within translation-invariant effective medium theories of this type, disorder generally suppresses the critical temperature $T_c$, with the exception of nonmagnetic impurities in an isotropic, $s$-wave paired superconductor,  where $T_c$ is impervious to disorder until the mean free path becomes of order an atomic spacing and localization effects set in.  The theory applies  equally well to unconventionally paired systems, where even nonmagnetic impurities are typically pair-breaking.  While it does not describe  $T_c$ suppression quantitatively in strongly coupled systems like cuprates, where Zn causes an initial suppression 2-3 times slower than the AG-rate\cite{exp_cuprates1,exp_cuprates2,exp_cuprates3,exp_cuprates4,exp_cuprates5}, still almost universally $T_c$ decreases upon addition of disorder.  

There are, however, a few special situations where this conclusion does not apply\cite{Palestini,Grest,Psaltakis,Chubukov_Vavilov_Fernandes16,Mishra2016,Feigel’man,Burmistrov,Mayoh,Yukalov1,Yukalov2,Mayoh2,Mariapaper,Kivelson,nunner,Dagotto,loh,scalettar,Mishra2008}.  We do not consider trivial $T_c$ enhancements, e.g. impurities that dope the system and thus change the Fermi surface, but rather physical effects of disorder itself not included in the AG approach for a simple BCS superconductor. For example, $T_c$ can  be enhanced by disorder if the superconductor is competing with another type of order, e.g. a density wave, which is more sensitive to disorder than the superconductor~\cite{Grest,Psaltakis,Chubukov_Vavilov_Fernandes16,Mishra2016}. Several authors have argued recently that $T_c$ can  be increased by disorder at levels where localization becomes important due to the multifractality of electronic wave functions\cite{Feigel’man,Burmistrov,Mayoh}. Related studies of $T_c$ enhancements exist also in the fields of granular and phase separated systems\cite{Yukalov1,Yukalov2,Mayoh2}. Finally we note a study where modulating the local density of states by disorder in several possible scenarios can yield an enhancement of $T_c$\cite{Mariapaper}.
 
Another class of studies have focused on effects of inhomogeneity in the pairing interaction itself without reference to any particular microscopic mechanism to create it\cite{Kivelson,nunner,Dagotto,loh,scalettar,Mishra2008}.  From these studies, it is known that systems with a modulated pair interaction have a $T_c$ that may be  enhanced  relative to a system with a homogeneous pairing interaction fixed  to the average in the modulated system\cite{Kivelson,scalettar}. Most theories of this type that rely on pairing inhomogeneity are somewhat idealized, however, since if the fluctuating pair interactions indeed arise from disorder, impurities or defects will inevitably create a concomitant screened Coulomb potential component that will tend to break pairs, particularly in unconventional superconductors.  

\begin{figure}[t]
 \begin{center}
 \includegraphics[width=7.5cm]{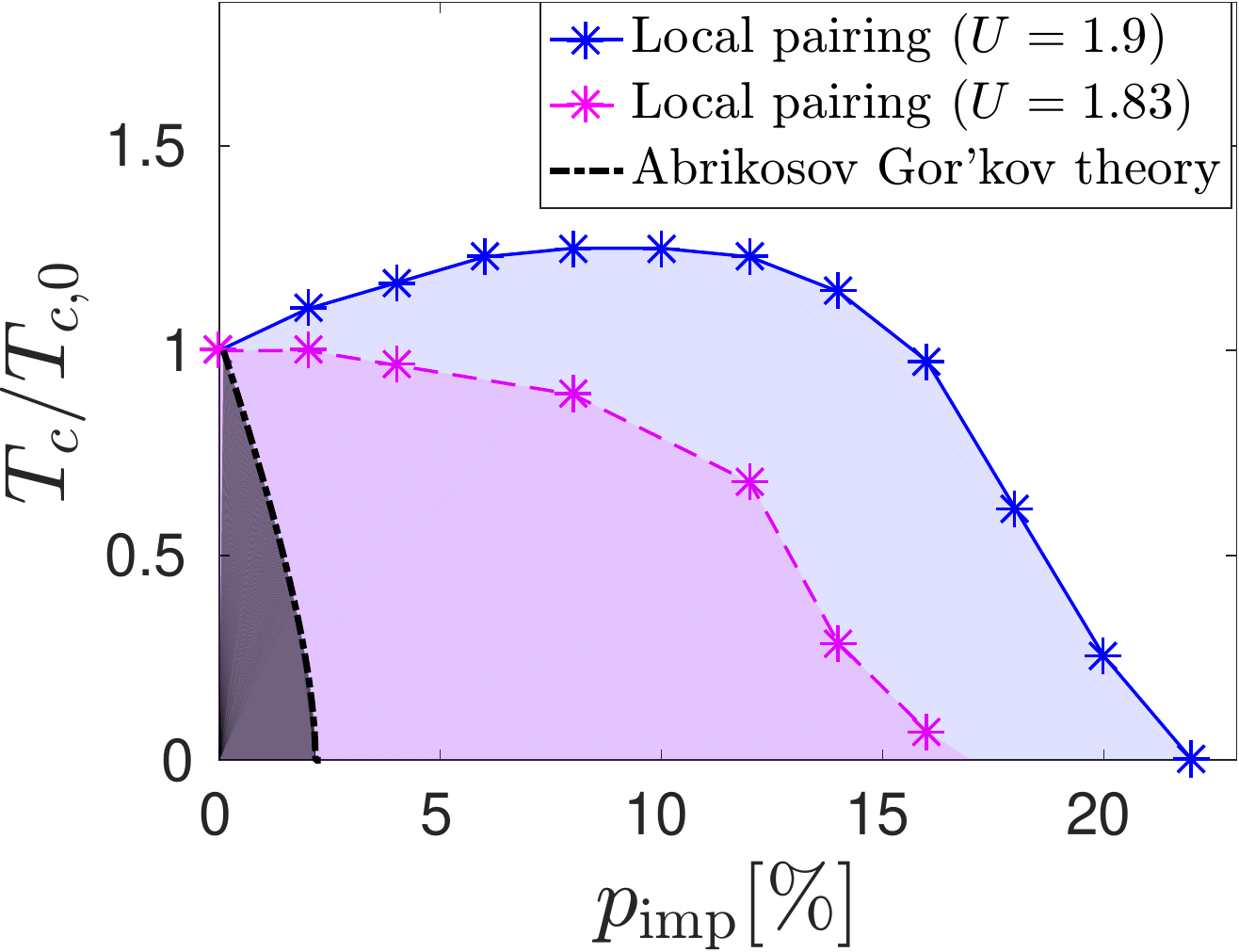}
 \end{center}
 \caption{Critical superconducting transition temperature
$T_c$ as a function of disorder concentration for nonmagnetic impurities  of  strength
$V_{imp}= 2$  in $d$-wave  superconductors of  Coulomb  interaction  strength $U= 1.9$  (blue  curve)  and
$U=1.83$  (magenta  curve).   Results  are  averaged  over  four different  impurity  configurations.   The  black  line  shows  the Abrikosov-Gor’kov result corresponding to the
$U= 1.9$ case.}
\label{fig:Vimp2}
\end{figure}

In this work, we propose a different mechanism for disorder-generated $T_c$-enhancements in unconventional superconductors. We study the effect of atomic scale defects on local spin fluctuations giving rise to $d$-wave pairing, but include pair-breaking effects through self-consistent studies of finite concentrations of disorder. From previous studies, it is known that a single nonmagnetic impurity softens spin fluctuations locally\cite{andersen07,graser10,mng13}, which favors $d$-wave pairing within a spin-fluctuation mediated scenario\cite{astrid12,foyevtsova2}. Note that the transfer of spectral weight is from typical normal state fluctuation energies of order $\sim t$ down to a fraction thereof; we do not treat dynamical pair-breaking effects known to occur when the fluctuations occur on the scale of $T_c$ itself\cite{MillisSachdevVarma}.  In terms of thermodynamics, however, such disorder-enhanced local pairing must compete with the inevitable pair-breaking effect of the impurities, and it is unclear which effect dominates $T_c$ for finite disorder concentrations $p_{\rm imp}$. As shown in Fig.~\ref{fig:Vimp2}, we find that the locally enhanced pairing scenario generally predicts significantly slower $T_c$-suppression rates, and can even in some circumstances support a remarkable disorder-generated $T_c$-enhancement. As seen from Fig.~\ref{fig:Vimp2}, this unusual behavior of $T_c$ is very different from that predicted by AG theory, which yields for a $d$-wave superconductor a rapid, monotonically decreasing $T_c$ with increasing disorder.

Specific to the one-band Hubbard model, we note the results of a recent dynamical cluster study of $d$-wave correlations finding a small initial enhancement of $T_c$ with $p_{\rm imp}$, and attributed it to an increase of the local exchange $J$ in a strong-coupling picture\cite{kemper09}. This study left unclear, however, under what circumstances  a system described by such a theory would exhibit conventional AG-like $T_c$ suppression with increasing $p_{\rm imp}$, and when  it will deviate strongly.  Under what circumstances can $T_c$ really be enhanced by the addition of disorder?  The present study was motivated in part by this theoretical question, and by recent electron irradiation experiments performed on FeSe\cite{Teknowijoyo}, which reported a 10\% rise in $T_c$ under circumstances that precluded an explanation in terms of doping or chemical pressure. Local pinning of spin fluctuations by irradiation-induced defects was one of the possible mechanisms discussed, but without reference to the possible pair-breaking effects that such defects could induce. 

\begin{figure*}[t]
 \begin{center}
 \includegraphics[angle=0,width=0.99\textwidth]{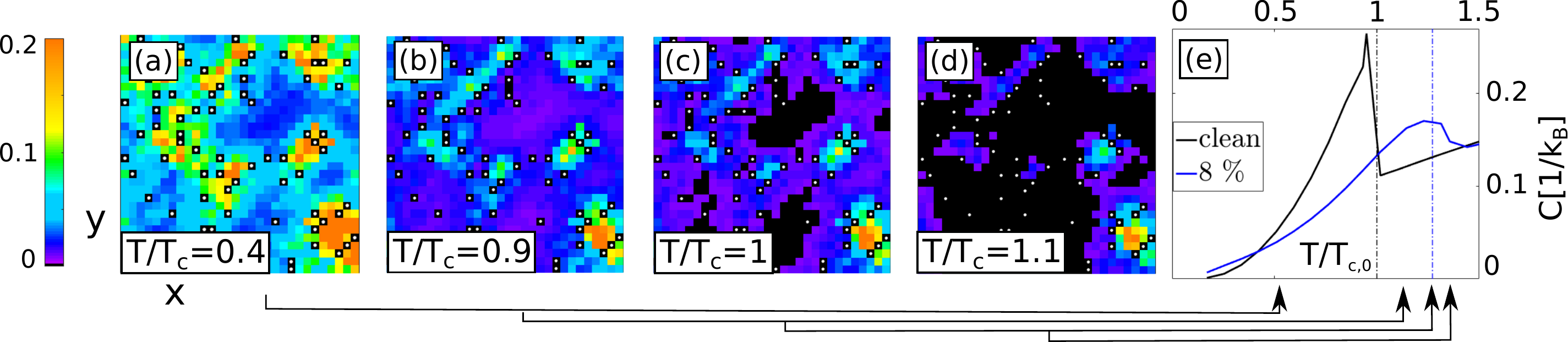}
 \end{center}
\caption{(a-d) Local gap maps below $T_c$ (a,b), at $T_c$ (c) and above $T_c$ (d) for a system of 8\% impurities with $V_{\rm imp}=2$. (e) Specific heat as a function of $T$ for the clean system (black line) and for 8\% disorder (blue line). The value of $T_c$ as defined by a finite gap exceeding $20\%$ of the homogeneous gap value at $T=0$ on 60\% of the sites is shown by the dashed lines in (e).}
\label{fig:specificheat}
\end{figure*}

{\it{Model and Method.}} The starting point is the one-band Hubbard model 
\begin{eqnarray}
H&=&-\sum_{i,j, \sigma}t_{i,j}c_{i\sigma}^{\dagger}c_{j\sigma}+ \sum_{i \sigma} U  n_{i\sigma} n_{i\bar\sigma} - \sum_{i \sigma} \mu  n_{i\sigma}   \nonumber \\
&&+ \sum_{i, i_{\rm imp}, \sigma} V_{\rm imp} \delta_{i, i_{\rm imp}}  n_{i\sigma},
\label{eq:Hhub}
\end{eqnarray}
with a concentration $p_{\rm imp}$ of nonmagnetic impurities of strength $V_{\rm imp}$ at random sites placed at positions $i_{\rm imp}$. The operator $c_{i\sigma}^{\dagger}$ refers to creation of an electron with spin $\sigma$ at lattice site $i$, and $n_{i\sigma}$ is the number operator of spin $\sigma$ particles at site $i$. The hopping elements $t_{i,j}$ include nearest neighbor (NN) $t=1$, and next-nearest neighbor (NNN) $t'=-0.3$, and the system is hole-doped by $x=0.15$, generating a standard Fermi surface relevant to cuprates. In the homogeneous case, an on-site repulsive Coulomb interaction $U$ gives rise to an effective attraction for superconductivity in the $d$-wave singlet channel as shown by weak-coupling spin-fluctuation theories \cite{Scalapino95,roemerPMpairing}, and in qualitative agreement with strong-coupling numerical studies\cite{maier05}. In the dirty case, however, $U$ modifies the charge and spin densities as well as the effective electron-electron interaction locally. To capture these effects, we first treat the Hubbard Hamiltonian at the mean-field level
\begin{eqnarray}
H_0&=&-\sum_{i,j, \sigma}t_{ij}c_{i\sigma}^{\dagger}c_{j\sigma}+ \sum_{i \sigma} (U \langle n_{i\sigma} \rangle - \mu ) n_{i\bar\sigma} \nonumber \\
&&+ \sum_{i,i_{\rm imp}, \sigma} V_{\rm imp} \delta_{i, i_{\rm imp}} n_{i\sigma},
\label{eq:H}
\end{eqnarray}
in order to determine the electronic densities self-consistently in the presence of the disorder. Given the self-consistent densities, the associated spatially modulated effective superconducting pairing arising from higher order interactions in $U$ is determined by~\cite{astrid12}
\begin{equation}
V^{\text{eff}}_{ij}=U+\frac{U^3\chi_0^2}{\hat{1}-U^2\chi_0^2}\Big|_{(i,j)}+\frac{U^2\chi_0}{\hat{1}-U\chi_0}\Big|_{(i,j)}.
\label{eq:Veff}
\end{equation}
The susceptibility in (\ref{eq:Veff}) is a real space matrix given by 
\begin{eqnarray}
\chi_{ij}^{\sigma \sigma'}\!=\!\sum_{m,n}u_{mi\sigma}u_{mj\sigma}u_{nj\sigma'}u_{ni\sigma'} \frac{f(E_{m\sigma})-f(E_{n\sigma'})}{E_{n\sigma'}-E_{m\sigma}+i\eta},
\end{eqnarray}
in terms of the eigenvectors $u_{m\sigma}$ and eigenvalues $E_{m\sigma}$ of Eq.(\ref{eq:H}). Thus, $u_{mi\sigma}$ denotes the value of the eigenfunction $u_{m\sigma}$ on site $i$. Note that, as is customary, the pairing interaction is assumed to be fully determined by the properties of the paramagnetic normal state.

After obtaining the effective self-consistent spin-fluctuation mediated pairing kernel in real space, the densities $\langle n_{i\sigma} \rangle$ and superconducting gap values $\Delta^s_{ij}$ are calculated via a second self-consistency loop from the full mean-field Hamiltonian given by
\begin{eqnarray}
H_{\rm SC}&=&H_0+\sum_{i,j}\left[\Delta^s_{ij}c_{i\up}^\dagger c_{j\down}^\dagger+\mbox{H.c.}\right]
\label{eq:Hdelta}
\end{eqnarray}
In the calculation of the singlet gaps 
\begin{equation}
\Delta^s_{ij}=-\frac{V^{\text{eff}}_{ij}}{2}
\sum_n [u_{ni}v_{nj} + u_{nj}v_{ni}]\tanh(E_n/2T),
\end{equation}
we account for superconducting links $\Delta_{i,i+\bf{\delta}}$, where $\pm \delta \in \{0,\hat{x},\hat{y},2\hat{x},2\hat{y},\hat{x}+\hat{y},\hat{x}-\hat{y}\}$. $\{E_n,u_n,v_n\}$ are the eigenvalues and eigenvectors resulting from diagonalization of Eq. (5). We refer to the above procedure as the "local pairing scenario". We find that in general, the NN links supporting $d$-wave superconductivity dominates, but higher order $d$-wave and subsidiary on-site order is induced in the vicinity of the impurities. We stress that the model contains only the free parameters $U$ and $V_{\rm imp}$. For the results below we fix $U=1.9$, and explore the dependence of $T_c$ on $V_{\rm imp}$ and $p_{\rm imp}$.

The results from the local pairing scenario are compared to standard AG theory of nonmagnetic impurities in unconventional (sign-changing) superconductors, where $T_c$ is obtained from the well known expression
\begin{equation}
\ln \Big(\frac{T_c}{T_{c,0}}\Big)=\Psi\Big(\frac{1}{2}\Big)-\Psi\Big(\frac{1}{2}+\frac{1}{4\pi T_c \tau}\Big).
\label{eq:AG}
\end{equation}
The normal state scattering rate in the $T$-matrix approximation is given by~\cite{hirschfeld86,SchmittRink86}
\begin{equation}
\frac{1}{\tau}=2\pi p_{\rm imp} \frac{V_{\rm imp}^2 N(0)}{1+(V_{\rm imp}N(0))^2},
\end{equation}
where $N(0)$ is the density of states at the Fermi level and $\Psi(x)$ refers to the digamma function.

{\it{Results.}}
For inhomogeneous systems, there are various definitions of $T_c$ that one might adopt. For example, one could define $T_c$ by the temperature at which the first island becomes superconducting upon cooling.
Instead, we adopt a more experimentally relevant definition: $T_c$ is the highest temperature where more than 60\% of the lattice sites possess a gap value that exceeds $20\%$ of $\Delta(0)$, where $\Delta(0)$ is the gap of the clean system at $T=0$ and $0.20\Delta(0)$ is of the order of the level spacing in our simulation, i.e. the bandwidth divided by system size $N^2$ with $N=30$. This rather conservative definition captures the situation where all superconducting sites of the 2D lattice percolate in the present case of randomly placed point-like disorder. Note our calculations are strictly at the level of inhomogeneous (BCS) mean field theory, and effects of fluctuations are therefore not included.  These fluctuations may be expected to suppress the mean field $T_c$ significantly in situations where the length scale of the inhomogeneity is larger than the coherence length\cite{Kivelson}, which is not the case here.

Local gap maps at temperatures both below and above $T_c$ are shown in Fig.~\ref{fig:specificheat}(a-d) for a system with 8\% impurities of strength $V_{\rm imp}=2$. We show the magnitude of the superconducting $d$-wave links calculated as $|\Delta_i|=\frac{1}{4}[\Delta_i(\hat x)-\Delta_i(\hat y) +\Delta_i(-\hat x)-\Delta_i(-\hat y)]$, where $\hat x$ ($\hat y$) denotes the unit vector along the $x$-axis ($y$-axis). At low $T$, large gap enhancements in the vicinity of the impurity sites are clearly visible as seen from Fig.~\ref{fig:specificheat}(a). Upon increasing temperature, the order is diminished and destroyed at sites farthest away from the impurities until eventually the superconducting regions become fully separated in space above $T_c$ as seen in Fig.~\ref{fig:specificheat}(d).

Due to the inhomogeneity of the superconducting phase, the thermodynamic response of the phase transition is smeared. We calculate the specific heat from the derivative of the entropy $C=T\partial S/\partial T$, where
\begin{equation}
S\!=\!\!-2 \sum_{E_n>0}f(E_n)\ln(f(E_n))\!+\!f(-E_n)\ln(f(-E_n)).
\end{equation}
The superconducting transition of the clean system is clearly manifested by a jump in the specific heat at $T_c$ as shown in Fig.~\ref{fig:specificheat}(e) by the black line. By contrast, in the dirty system with 8\% disorder, a broad peak marks the transition at a temperature that agrees well with the definition of $T_c$ stated above~\cite{andersen06}.

In Fig.~\ref{fig:Vimp2} we show the full evolution of $T_c$ versus $p_{\rm imp}$ for the case with $V_{\rm imp}=2$. The $T_c$-enhancement is clearly visible in an extended range of disorder concentrations in the case with $U=1.9$. For weaker $U$, $T_c$ is suppressed for all $p_{\rm imp}$ but still exhibits a large critical impurity concentration. In fact, within the local pairing scenario the superconductor is much more robust to impurities than predicted by AG theory, easily supporting a superconducting state to an order of magnitude more disorder as seen from Fig.~\ref{fig:Vimp2}. Figure~\ref{fig:Vimp2} thus demonstrates that indeed the local pairing enhancements caused by the softened spin fluctuations can overcome the inevitable pair-breaking for a significant range of $p_{\rm imp}$. A similar study for attractive impurities~\cite{suppl} reveals that the $T_c$-suppression rate remains weaker than prescribed by AG-theory, but no disorder-generated $T_c$-enhancement exists in the case of $V_{\rm imp}<0$ for the cuprate-like band structure studied here.

\begin{figure}[t]
 \begin{center}
 \includegraphics[width=9.cm]{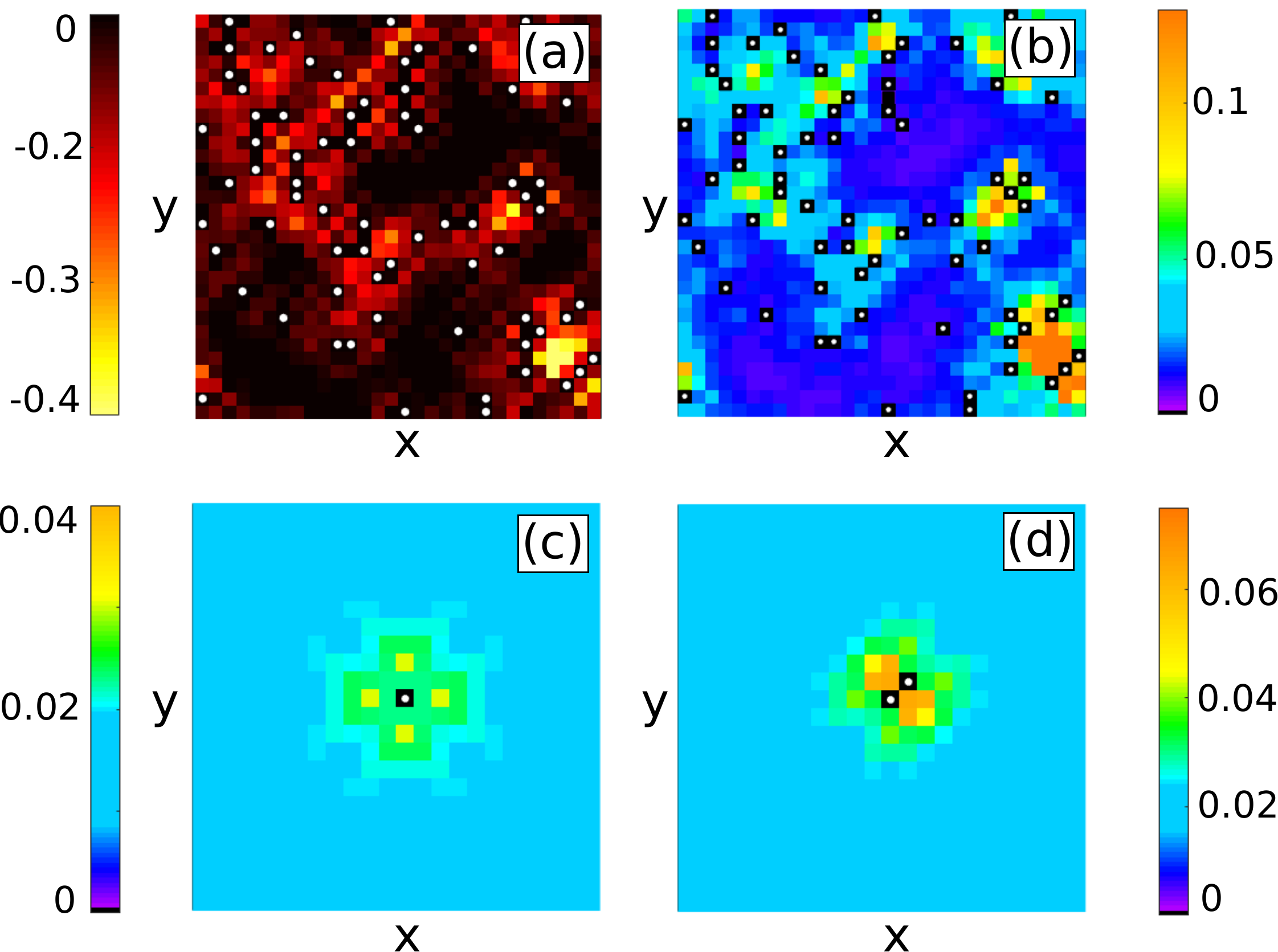}
 \end{center}
\caption{(a) Real space map of the increase in NN pairing attraction above the pairing strength of the clean system $V^{\rm eff}-V^{\rm eff}_0$ at $T=1.2T_{c,0}$, where $T_{c,0}$ is the critical temperature of the clean system. The system contains 8\% impurities of strength $V_{\rm imp}=2$ (white dots). (b) The resulting local $d$-wave gap map for the same system as in (a). Black sites have $\Delta_i < 0.2\Delta(0)$. (c,d) Local gap at $T=0.7T_{c,0}$ around a single impurity (c) and two impurities in diagonal-dimer formation (d) of strength V$_{\rm imp}=2$. Note the difference in color scale. 
}
\label{fig:vparenhancement}
\end{figure} 

In order to understand the origin of the $T_c$ enhancement of Fig.~\ref{fig:Vimp2}, we show in Fig.~\ref{fig:vparenhancement}(a) the increase in NN attraction $\frac{1}{4}[V^{\rm eff}_{i,i+\hat{x}}+V^{\rm eff}_{i,i+\hat{y}}+V^{\rm eff}_{i,i-\hat{x}}+V^{\rm eff}_{i,i-\hat{y}}]$ for a system of 8\% impurities, still with $V_{\rm imp}=2$. We calculate the pairing of the dirty system $V^{\rm eff}(T)$ at $T=1.2T_{c,0}$, where $T_{c,0}$ is the critical temperature of the clean system and subtract the NN attraction in the pure case $V^{\rm eff}_0(T_{c,0})$. We stress that the attraction  in the dark regions of  Fig.~\ref{fig:vparenhancement}(a) is not in itself sufficient to support superconductivity (since $T>T_{c,0}$).   Nevertheless, the system displays a non-zero $d$-wave gap in these regions, as seen from Fig.~\ref{fig:vparenhancement}(b), due to proximity coupling to the regions of enhanced pairing, which thereby boost the superconducting condensate of the entire system. Such local regions favorable to pairing can be understood from certain advantageous clustering of impurities, illustrated in Fig.~\ref{fig:vparenhancement}(c) and (d). For example, a constructive interference of two impurities forming diagonal dimers lead to gap enhancements of $\sim 200$\% with 6 sites involved, as compared to the $\sim 50$\% enhancement effect of four sites around a single impurity. Diagonal structures of more than two impurities are even more advantageous and systems with such structures lead to an even larger increase in local pairing. 

\begin{figure}[t]
 \begin{center}
 \includegraphics[width=8.cm]{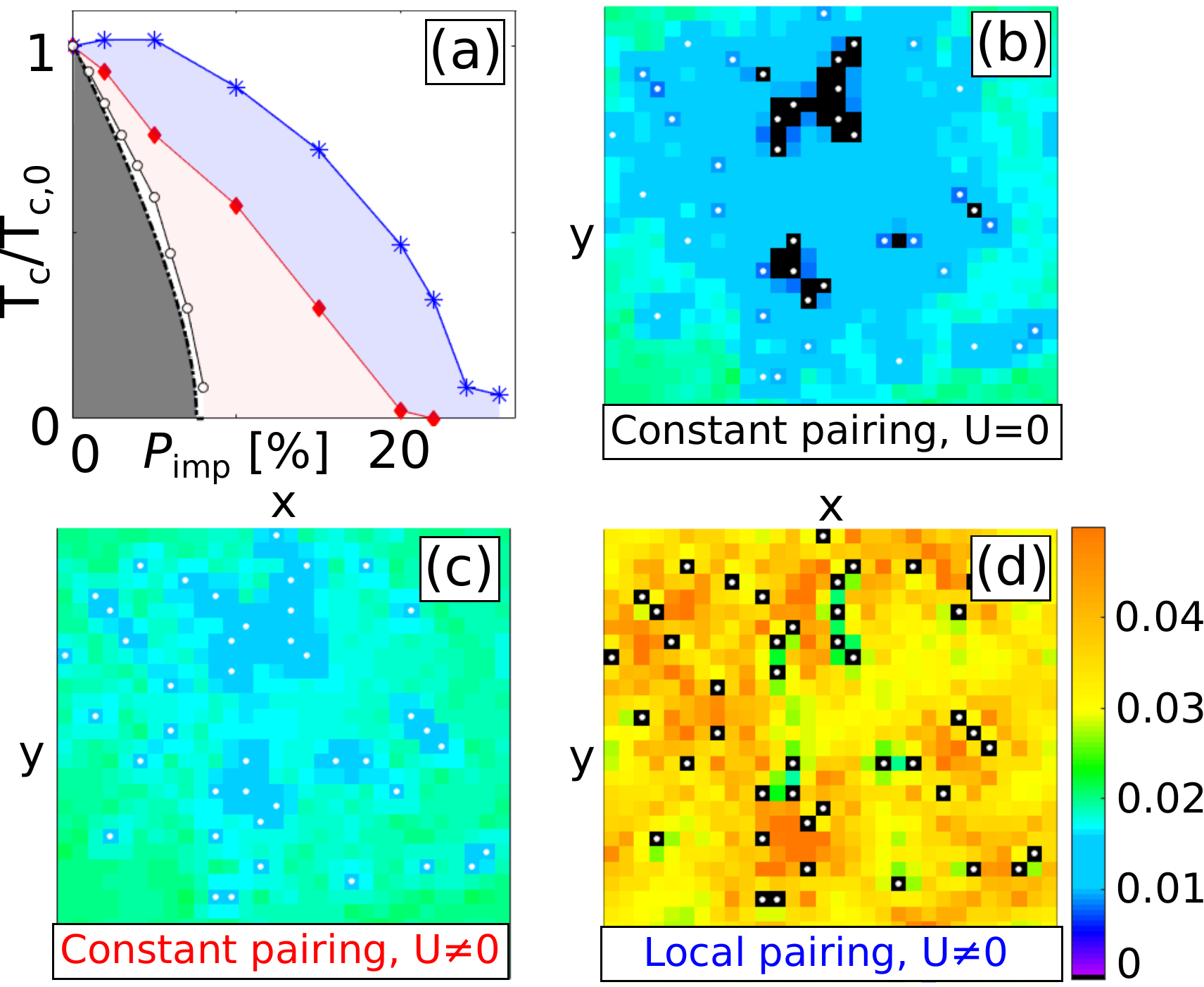}
 \end{center}
\caption{(a) Suppression of $T_c$ versus nonmagnetic disorder concentration, $V_{\rm imp}=1$. The dashed line refers to AG theory. The open circles correspond to a real space calculation with $U=0$ and constant pairing, roughly confirming the AG result, as expected. The red (blue) curve shows the $T_c$-suppression for $U=1.9$, and constant pairing (inhomogeneous local pairing). (b-d) Magnitude of the local $d$-wave gap in a system with 5\% disorder at $T=0.7T_{c,0}$. Impurity positions are marked by white dots. The gap maps correspond to the cases of $U=0$, constant pairing (b); $U=1.9$, constant pairing (c); and $U=1.9$, local pairing (d).}
\label{fig:vimp1_Tcsuppression}
\end{figure}

In Fig.~\ref{fig:vimp1_Tcsuppression}(a) we show the results of the $T_c$-suppression rate for the case with a weaker impurity potential $V_{\rm imp}=1$. As expected, weaker scatterers raise the critical disorder concentration. However, it is found that 1) there remains a substantial difference between the AG result and the local pairing scenario, and 2) the $T_c$ enhancement is nearly eliminated. There are two reasons for property 1); correlation-induced screening\cite{Andersen08,Garg08,kemper09,ghosal00,chakraborty14,tang16,chakraborty17}, and local pairing enhancements. By performing the real-space calculation for the case $U=0$, while including a constant nearest-neighbor attraction, one almost quantitatively obtains the AG result, despite the local suppressions of the gap. However, as an instructive intermediate step we have calculated the $T_c$-suppression when $U\neq0$, but without local pairing modulations, as shown by the red curve in Fig.~\ref{fig:vimp1_Tcsuppression}(a). A comparison of gap maps in Fig.~\ref{fig:vimp1_Tcsuppression}(b) and in Fig.~\ref{fig:vimp1_Tcsuppression}(c) reveals a less modulated gap for the case $U\neq0$ than for $U=0$. This correlation-induced screening arises from the induced density modulations at the impurity site as seen by rewriting the density mean-field term as
$\sum_{i \sigma} U \langle n_{i\sigma} \rangle n_{i\bar\sigma} = \sum_{i \sigma} U [\Delta n_{i\sigma} n_{i\bar\sigma}+\frac{n_0}{2}(n_{i\sigma}+n_{i\bar\sigma})]$, where $\Delta n_{i}=\langle n_{i}\rangle - n_0$, and $n_0$ denotes the density of the clean system. The presence of a local repulsive potential repels electrons from the impurity site creating a $\Delta n_{\rm imp}=\langle n_{\rm imp}\rangle - n_0 <0$ . This reduces the effective impurity potential $[V_{\rm imp}+U\Delta n_{\rm imp}]$, an effect most relevant to weak impurity potentials, and reduces their $T_c$-suppression rate. The opposite effect happens for magnetic impurities, which are anti-screened by $U$\cite{mng16}. The $T_c$-suppression rate is further decreased when the electronic correlations are included also in the effective pairing interaction for  the inhomogeneous system, as seen from Fig.~\ref{fig:vimp1_Tcsuppression}(a), and the comparative gap map in Fig.~\ref{fig:vimp1_Tcsuppression}(d). We note that the value of $U$ at the impurity sites affects the screening effect, but does not modify $T_c$ in the local pairing approach since the pairing enhancement is not occurring at the impurity sites, but in their vicinity.

Regarding point 2) above, stronger individual impurities of $V_{\rm imp}\simeq 2$ lead to larger local pairing on neighboring sites compared to $V_{imp} \leq 1$. At small to moderate concentrations $p_{\rm imp}$, stronger impurities are therefore more beneficial for the global $T_c$. However, a larger impurity potential is more pair-breaking, and therefore at large $p_{\rm imp}$ the pair-breaking effect becomes dominant in agreement with the decreasing critical impurity concentration for larger impurity potentials. In the unitary limit the density is fully suppressed at the impurity sites, and $T_c$ is independent of $V_{\rm imp}$~\cite{suppl}. In this limit, the pair-breaking effect  dominates at all impurity concentrations and $T_c$ is determined by $p_{\rm imp}$ alone.

In conclusion, we have shown how atomic-scale disorder generates highly inhomogeneous effective pairing interactions within a spin-fluctuation pairing scenario. This results in a superconducting phase with local regions of large gap enhancements compared to the homogeneous system, and makes the superconductor much more robust to disorder, in some cases  enhancing $T_c$ of the disordered system. The mechanism described in this work is enhanced for larger impurity potentials, and by the proximity of the system to a magnetic instability.  It is a likely explanation for the well-known slower decrease of $T_c$ with disorder in cuprates relative to that anticipated from AG theory\cite{exp_cuprates1,exp_cuprates2,exp_cuprates3,exp_cuprates4,exp_cuprates5}, and may also be related to a recently observed increase of $T_c$ with electron irradiation in FeSe\cite{Teknowijoyo}.

We acknowledge useful discussions with D. Chakraborty, J. Dodaro, M. N. Gastiasoro, A. Ghosal and S. A. Kivelson. B.M.A. and A.T.R. acknowledge support from a Lundbeckfond fellowship (Grant No. A9318).  P.J.H. was supported by NSF Grant No. DMR-1407502, and is grateful to Stanford Institute for Materials and Energy Sciences and Stanford
Institute for Theoretical Physics for support during the preparation of this manuscript.


\newpage
\begin{center}
\bf{Supplementary Material: "Raising the critical temperature by disorder in unconventional superconductors mediated by spin fluctuations" }

\end{center}
\begin{center}
 \small
Astrid T. R\o mer,$^1$ P. J. Hirschfeld,$^2$ Brian M. Andersen$^1$\\
$^1$Niels Bohr Institute, University of Copenhagen, Juliane Maries Vej 30, DK-2100 Copenhagen,
Denmark\\
$^2$Department of Physics, University of Florida, Gainesville, Florida 32611, USA

\end{center}




In this supplementary material we address the local pairing scenario when impurities are described by attractive potentials and compare the results to the local pairing for repulsive impurities. Furthermore, we explore the gap and $T_c$ dependence on the impurity strength and consider the limit of very strong repulsive impurities.
Lastly, we elaborate the discussion of the screening of impurities by electronic density feedback which occurs in the case of repulsive as well as attractive impurities.

\section{Local pairing for attractive impurities}
\begin{figure}[b]
 \begin{center}
 \includegraphics[angle=0,width=0.45\textwidth]{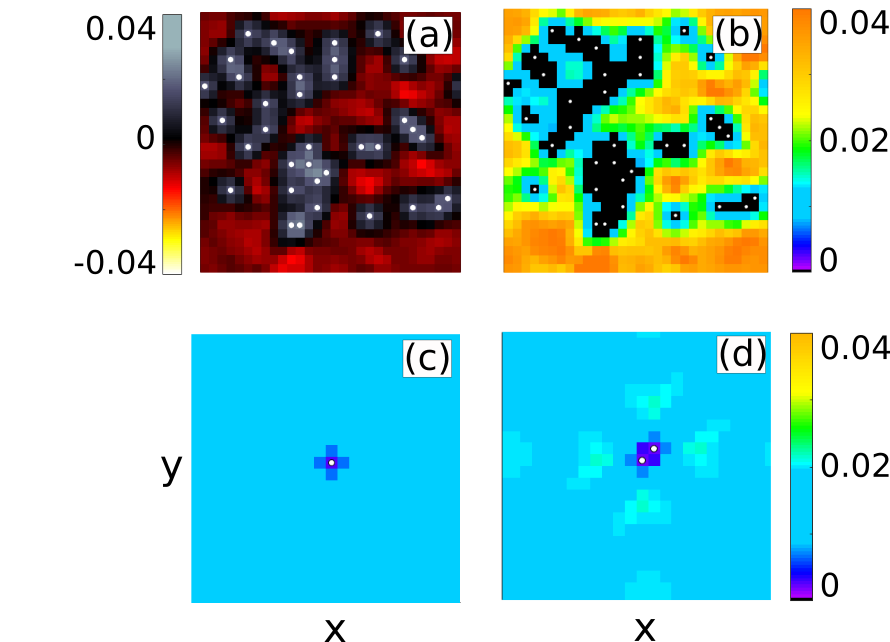}
 \end{center}
\caption{(a) Real space map ($30\times 30$ lattice sites) of the difference in nearest-neighbor interaction compared to the pairing of the clean system at $T/T_{c,0} =0.8$ for 4 \% attractive impurities with $V_{\rm imp}=-2$ and $U=1.9$. $T_{c,0}$ denotes the critical temperature of the clean system. (b) Local $d$-wave gap map for the same system as (a) . (c,d) Local gap map around a single and dimer attractive impurity with $V_{\rm imp}=-2$ at $T/T_{c,0} =0.7$.  The average density is $\langle n \rangle =0.85$ and $t'=-0.3$.}
\label{fig:Vimpm2}
\end{figure}

In the Abrikosov-Gor'kov approach to dirty superconductors, the suppression rate of $T_c$ is insensitive to the sign of the impurity potential. Also screening of the impurity potential arising from density modulations is roughly independent of whether the impurity is attractive or repulsive as we show in Sec.~\ref{sec:screen}. On the contrary, insensitivity to the sign of the impurity potential does not occur in the full model where the pairing interaction is a result of the local antiferromagnetic fluctuations. Due to the electron-hole asymmetry of the electronic band relevant to the cuprate system, antiferromagnetic fluctuations become stronger in the vicinity of repulsive impurities, as compared to the vicinity of attractive impurities.
\begin{figure}[b]
 \begin{center}
 \includegraphics[angle=0,width=0.45\textwidth]{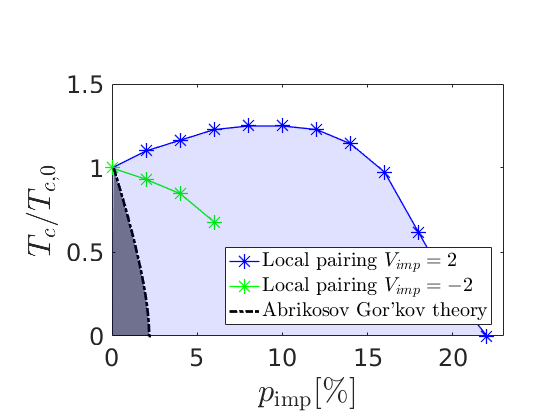}
 \end{center}
\caption{The evolution of $T_c$ upon increasing impurity concentration for a system with repulsive ($V_{\rm imp}=2$) and attractive ($V_{\rm imp}=-2$) impurities in a system with $U=1.9$. The average density is $\langle n \rangle =0.85$ and $t'=-0.3$. In case of attractive impurities we explored impurity concentrations $p_{\rm imp} \leq 6$ \% to investigate the presence or absence of a $T_c$ increase at low concentrations. Attractive impurities do not give
rise to a $T_c$ increase, but cause
suppression of the $T_c$ decrease rate compared to the Abrikosov-Gor'kov result. For moderate concentrations of repulsive impurities there is an enhancement of $T_c$. }
\label{fig:Vimpm2_Tccurve}
\end{figure}

In the case of many attractive impurities, the local pairing interaction will suffer a complete destruction at the impurity site as well as in its vicinity and only very small pairing enhancements are found in regions far from the impurities as shown in Fig.~\ref{fig:Vimpm2} (a). This results in a destruction of the superconducting gap with the exception of regions where the pairing is not affected or weakly enhanced, see Fig.~\ref{fig:Vimpm2} (b).
This result is very different from the result of repulsive impurities of strength $V_{\rm imp}=2$ considered in Fig. 3 of the main paper. 
For further comparison we show in
Fig.~\ref{fig:Vimpm2} (c,d) the local gaps around an attractive single impurity of $V_{\rm imp}=-2$  and dimer impurity potential of the same strength. 
There is no $d$-wave enhancement of the single impurity and only a very slight enhancement of the dimer structure further away from the impurities. 
The complete destruction of the gap at the impurity site that was observed for repulsive impurities is weakened and becomes only a gap depression in the case of single and dimer attractive impurities, as seen from Fig.~\ref{fig:Vimpm2} (c,d) at the impurity sites.

The difference between the system's response to repulsive and attractive impurities can be visualized by calculating the local spin response by a partial Fourier transformation in relative momentum transfer  $\chi_0({\bf R},Q)$ for $Q=(\pi,\pi)$~\cite{astrid12}. Note that the presence of the impurity potentials in the normal-state Hubbard Hamiltonian $H_0$ (Eq. (2) of the main paper) ensures that the impurity effect is present at the level of the bare susceptibility. The system is driven towards a local antiferromagnetic Stoner instability 
$URe\chi_0({\bf R},Q) \to 1$ for the lattice sites surrounding a repulsive impurity, whereas this effect is absent when the impurity is attractive.  
Therefore the prevailing effect of attractive impurities in the cuprate band is a local pairing decrease. The small pairing increase in some regions as visualized in Fig.~\ref{fig:Vimpm2} (a) it is not sufficient to overcome the pair-breaking response of superconductivity to the impurities and the critical temperature decreases in the disordered system for all impurity concentrations, as shown in Fig.~\ref{fig:Vimpm2_Tccurve}. The $T_c$ suppression rate is still slower than prescribed by AG-theory, however.

\begin{figure}[b]
 \begin{center}
 \includegraphics[angle=0,width=0.5\textwidth]{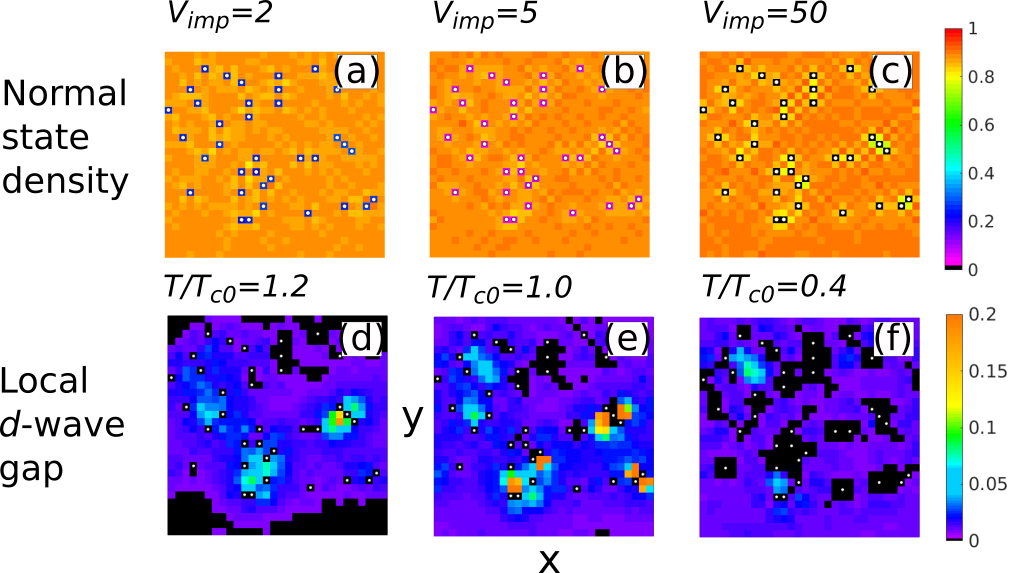}
 \end{center}
\caption{Real space plots on $30\times 30$ lattices of the normal state density (a-c) and local $d$-wave gap at $T=T_c$ of the disordered system (d-f) for 4 \% impurities of strength $V_{\rm imp}=2, 5, 50$ in a system with $U=1.9$. Sites with $\Delta _i<0.2\Delta(0)$ (where $\Delta(0)$ is the gap of the clean system at $T=0$) are black. $T_{c0}$ denotes the critical temperature of the clean system.  The average density is $\langle n \rangle = 0.85$ and $t'=-0.3$. }
\label{fig:unlim_2Dplots}
\end{figure}

\section{The unitary limit}
Within Abrikosov-Gor'kov theory the scattering rate in Eq. (8) of the main paper becomes $1/\tau=2\pi p_{\rm imp}/N(0)$ in the unitary limit of $V_{\rm imp} \to \infty$. In this limit, the scattering rate is thus independent of $V_{\rm imp}$ and the critical temperature will not depend on variations of  the impurity strength in this regime. 

For a fixed impurity concentration $p_{\rm imp}$ in the dilute limit,  AG theory predicts  an initial decrease of $T_c$ as function of $V_{\rm imp}$ followed by a saturation  in the unitary limit.  
\begin{figure}[b]
 \begin{center}
 
 \includegraphics[angle=0,width=0.37\textwidth]{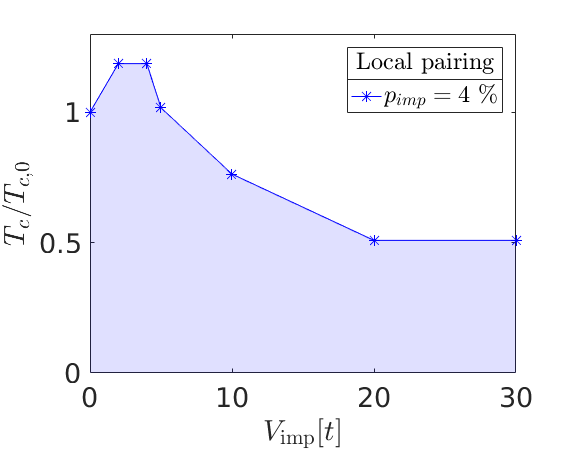}
 \end{center}
\caption{$T_c$ versus impurity strength for repulsive potential impurities and $U=1.9$. For moderate impurity strengths ($V_{\rm imp}\leq 5$) an increase in $T_c$ is observed. In the unitary limit ($V_{\rm imp}\geq 20$) $T_c$  is independent of the potential strength due to a complete depression of the local density at the impurity sites. In this regime, the pair-breaking effect dominates. }
\label{fig:unlim}
\end{figure}

In the local pairing model, the initial behavior is qualitatively different. The electronic density distribution and spin fluctuation physics change with the 
strength of the impurities and for moderate repulsive impurity potentials this leads to a $T_c$ increase, as shown in the main paper as well as in Fig.~\ref{fig:Vimpm2_Tccurve}. An increase in the impurity potential is considered in Fig.~\ref{fig:unlim_2Dplots} where the normal state density and local superconducting gap are depicted in the case of 4\% moderate, strong and extremely strong impurities. For moderate impurity strength $V_{\rm imp}=2$, the gap is suppressed at the impurity sites and additional gap closings are first encountered in regions far from the impurities upon increasing temperature. On the contrary, regions close to the impurities have a finite gap, even for temperatures very close to $T_c$ ($>T_{c,0}$) of the dirty system as seen from Fig.~\ref{fig:unlim_2Dplots} (d). We note particularly that  the largest gap values appear close to local dimer-like formations of impurities, as also discussed in the main paper. 
Towards the unitary limit, e.g. $V_{\rm imp}=50$, the  impurity sites will suffer a complete electron density suppression, see Fig.~\ref{fig:unlim_2Dplots} (c).
In this limit, the pair-breaking effect dominates the regions surrounding the impurities as well and at the critical temperature the regions around the impurities experience gap closings as shown in Fig.~\ref{fig:unlim_2Dplots} (f).
For an intermediate potential strength of $V_{\rm imp}=5$, Fig.~\ref{fig:unlim_2Dplots} (b,e), there is a compromise between even larger local gap enhancements compared to $V_{\rm imp}=2$ and an increasing pair-breaking effect for neighbor sites of the impurities.

Despite the difference between the local pairing model and Abrikosov-Gor'kov theory, both approaches actually find that the critical temperature is independent of the impurity strength in the unitary limit. In the local pairing picture, this is due to the fact that the normal state density will display the same local pattern for all impurity strengths above a certain threshold value. Thus, even in the local pairing model, the critical temperature will become independent of $V_{\rm imp}$ in the unitary limit, which is shown in Fig.~\ref{fig:unlim}. In this limit the pair-breaking effect of the impurities will dominate and the critical temperature will be suppressed compared to the the critical temperature of the clean system $T_{c,0}$. The saturation limit will depend on the impurity concentration. 
\begin{figure*}[b!]
 \centering
 \includegraphics[angle=0,width=0.8\textwidth]{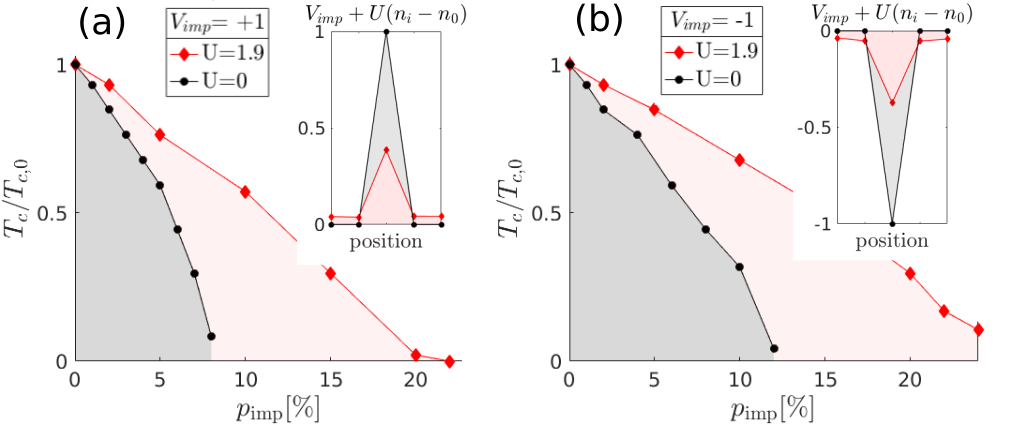}
 \caption{Screening effect due to correlations for (a) repulsive and (b) attractive impurities. The black circles display the critical temperature for a constant pairing interaction with no screening ($U=0$). The red diamonds show the critical temperature for the same pairing interaction including the density mean-field correlations in the self-consistent calculation (i.e. $U\neq 0$). The insets show the local density at the impurity site for repulsive and attractive impurity of $V_{\rm imp}=\pm 1$, respectively.}
\label{fig:screen}
\end{figure*}

\section{Coulomb screening of attractive and repulsive potential impurities}
\label{sec:screen}
In this section we elaborate the discussion on how the bare impurity potential becomes screened due to feedback from density modulations. This effect gives rise to a much slower $T_c$ decrease rate than the corresponding Abrikosov-Gor'kov calculation. Screening of potential impurities is equally effective for repulsive and attractive impurities. When the nearest-neighbor attraction is finite at the impurity sites this will lead to a screening effect in the superconducting channel, i.e. a smoothing of the superconducting gap and suppression of the $T_c$ decrease rate. The screening effect on the gap is clearly observed when a constant pairing interaction is invoked \cite{Andersen08,Garg08} and will always be present if the impurity sites possess a finite pairing amplitude.

The mean-field decoupling of the bare Hubbard interaction $U$ accounts for feedback effects due to spin-density modulations,$\langle n_{i\up} \rangle$  and $\langle n_{i\down} \rangle$. This decoupling allows for the realization of an antiferromagnetic spin-density wave in the model when the Hubbard $U$ is above a critical interaction strength $U_c$ which depends on the band and filling. We operate below $U_c$ and do not consider coexistence of static antiferromagnetism and superconductivity.
In the disordered system it is possible to have density modulations {\it without} any difference between the densities of spin-up and spin-down electrons. Thus, there is a local feedback effect from the density modulations even for $U<U_c$, i.e. in the absence of static antiferromagnetic order. The significance of the mean-field interaction term $U\langle n_{i \sigma} \rangle n_{i \overline \sigma} $ is to account for density modulation feedback effects.
The presence of an impurity causes a local suppression or enhancement of the local electronic density depending on the sign (+/-) of the impurity potential. 
A repulsive impurity potential $V_{\rm imp}>0$ will repel electrons from the impurity site and this will build up a local negative density fluctuation residing at the impurity site, $\Delta n_{\rm imp}=\langle n_{\rm imp}\rangle-n_0 < 0$ where $n_0$ is the average density of the system, which is kept fixed. If instead an attractive impurity potential is invoked, electrons will be attracted towards the impurity site, i.e. $\Delta n_{\rm imp}>0$. 

The feedback effect from density modulations can be described as a screening effect in the following way:  Irrespective of the sign of the impurity potential we will have $|V_{\rm imp}+U\Delta n_{\rm imp} |< |V_{\rm imp}|$ because $\Delta n_{\rm imp}$ is negative(positive) for $V_{\rm imp}$ positive(negative). This means that the local density modulation will diminish the bare impurity potential in the Hartree approximation, but only if electronic interactions are allowed through a non-zero Hubbard $U$. The correlation effect is absent for $U=0$. To illustrate the case for both repulsive and attractive impurities we plot in the insets of  Fig.~\ref{fig:screen} two cuts through an impurity site to compare the bare potential (black curve) with the screened potential (red curve). The screened potentials visualized in Fig.~\ref{fig:screen} lead to a much weaker pair-breaking effect. Therefore, the inclusion of density modulations by a non-zero Hubbard $U$ leads to a much slower $T_c$ decrease as a function of impurity concentration. This is shown in Fig.~\ref{fig:screen} for both types of impurities.

As a final remark it should be mentioned that the local value of the Coulomb interaction {\it at} the impurity sites does not affect $T_c$ in the local pairing scenario with repulsive potential impurities. The enhancement effect of spin-fluctuation mediated pairing is not occurring at the impurity sites, but in their vicinity. Therefore, the Hubbard $U$ {\it at} the impurity site is in fact unimportant for the resulting superconducting gap of the local pairing calculation. In particular,   the $T_c$ enhancement effect is independent of
whether we associate the impurity sites with a local $U$ or set $U=0$ at all impurity sites.  
\vskip .2cm



\begin{thebibliography}{}

\bibitem{Anderson59}
P. W. Anderson, {\it Theory of dirty superconductors}, J. Phys. Chem. Solids {\bf 11}, 26 (1959).

\bibitem{ag} A. A. Abrikosov and L. P. Gorkov, Zh. Eksp. Teor. Fiz. {\bf 39}, 1781 (1960). A. A. Abrikosov and L. P. Gorkov,
{\it Contribution to the theory of superconducting alloys with paramagnetic impurities},
Sov. Phys. JETP {\bf 12}, 1243 (1961).

%
\bibitem{exp_cuprates1} D. N. Basov, A. V. Puchkov, R. A. Hughes, T. Strach, J. Preston, T. Timusk, D. A. Bonn, R. Liang, and W. N. Hardy, {\it Disorder and superconducting-state conductivity of single crystals of YBa$_2$Cu$_3$O$_{6.95}$}, Phys. Rev. B {\bf 49}, 12 165 (1994).
%
\bibitem{exp_cuprates2} E. R. Ulm, J.-T. Kim, T. R. Lemberger, S. R. Foltyn, and X. Wu, {\it Magnetic penetration depth in Ni- and Zn-doped 
YBa$_2$(Cu$_{1−x}$M$_x$)$_3$O$_7$ films}, Phys. Rev. B {\bf 51}, 9193 (1995).
%
\bibitem{exp_cuprates3} B. Nachumi, A. Keren, K. Kojima, M. Larkin, G. M. Luke, J. Merrin, O. Tchernysh\"{o}v, Y. J. Uemura, N. Ichikawa, M. Goto, and S. Uchida, {\it Muon Spin Relaxation Studies of Zn-Substitution Effects in High-T$_c$ Cuprate Superconductors}, Phys. Rev. Lett. {\bf 77}, 5421 (1996).
%
\bibitem{exp_cuprates4} S. K. Tolpygo, J.-Y. Lin, M. Gurvitch, S. Y. Hou, and J. M. Phillips, {\it Universal T$_c$ suppression by in-plane defects in high-temperature superconductors: Implications for pairing symmetry}, Phys. Rev. B {\bf 53}, 12454 (1996).
%
\bibitem{exp_cuprates5} C. Bernhard, J. L. Tallon, C. Bucci, R. De Renzi, G. Guidi, G. V. M. Williams, and Ch. Niedermayer, {\it Suppression of the Superconducting Condensate in the High- T$_c$ Cuprates by Zn Substitution and Overdoping: Evidence for an Unconventional Pairing State}, Phys. Rev. Lett. {\bf 77}, 2304 (1996).

\bibitem{Palestini} F. Palestini and G. C. Strinati, {\it Systematic investigation of the effects of disorder at the lowest order throughout the BCS-BEC crossover}, 
Phys. Rev. B {\bf 88}, 174504 (2013).

\bibitem{Grest} G. S. Grest, K. Levin, and M. J. Nass, {\it Impurity and fluctuation effects in charge-density-wave superconductors}, Phys. Rev. B {\bf 25}, 4562 (1982).

\bibitem{Psaltakis} G. C. Psaltakis, {\it Non-magnetic impurity effects in charge-density-wave superconductors}, J. Phys. C: Solid St. Phys. {\bf 17}, 2145 (1984).

\bibitem{Chubukov_Vavilov_Fernandes16} R. M. Fernandes, M. G. Vavilov and A. V. Chubukov, {\it Enhancement of T$_c$ by disorder in underdoped iron pnictide superconductors}, Phys. Rev. B {\bf 85}, 140512 (2012).

\bibitem{Mishra2016} V. Mishra and P. J. Hirschfeld, {\it Effect of disorder on the competition between nematic and superconducting order in FeSe}, New J. Phys. {\bf 18}, 103001 (2016).

\bibitem{Feigel’man} M. V. Feigel’man, L. B. Ioffe, V. E. Kravtsov, and E. A. Yuzbashyan, {\it Eigenfunction Fractality and Pseudogap State near the Superconductor-Insulator Transition}, Phys. Rev. Lett. {\bf 98}, 027001 (2007).

\bibitem{Burmistrov} I. S. Burmistrov, I. V. Gornyi, and A. D. Mirlin, {\it Enhancement of the Critical Temperature of Superconductors by Anderson Localization}, 
Phys. Rev. Lett. {\bf 108}, 017002 (2012).

\bibitem{Mayoh} J. Mayoh and A. M. Garc\'{i}a-Garc\'{i}a, {\it Global critical temperature in disordered superconductors with weak multifractality}, Phys. Rev. B {\bf 92}, 174526 (2015).

\bibitem{Yukalov1} A. J. Coleman, E. P. Yukalova, V. I. Yukalov, {\it Superconductors with mesoscopic phase separation}, Physics C {\bf 243}, 76 (1995).

\bibitem{Yukalov2} V. I. Yukalov and E. P. Yukalova, {\it Mesoscopic phase separation in anisotropic superconductors}, Phys. Rev. B {\bf 70}, 224516 (2004).

\bibitem{Mayoh2} J. Mayoh and A. M. Garc\'{i}a-Garc\'{i}a, {\it Strong enhancement of bulk superconductivity by engineered nanogranularity}, Phys. Rev. B {\bf 90}, 134513 (2014).

\bibitem{Mariapaper} M. N. Gastiasoro and B. M. Andersen 
{\it Enhancing Superconductivity by Disorder}, arXiv:1712.02656

\bibitem{Kivelson} I. Martin, D. Podolsky and S. A. Kivelson, {\it Enhancement of superconductivity by local inhomogeneities}, Phys. Rev. B {\bf 72}, 060502 (2005); E. Arrigoni and S. A. Kivelson, {\it Optimal inhomogeneity for superconductivity}, Phys. Rev. B {\bf 68}, 180503 (2003).

\bibitem{nunner} T. S. Nunner, B. M. Andersen, A. Melikyan, and P. J. Hirschfeld, {\it Dopant-Modulated Pair Interaction in Cuprate Superconductors}, Phys. Rev. Lett. {\bf 95}, 177003 (2005).

\bibitem{Dagotto} G. Alvarez, M. Mayr, A. Moreo, and E. Dagotto, {\it Areas of superconductivity and giant proximity effects in underdoped cuprates}, Phys. Rev. B \textbf{71}, 014514 (2005).

\bibitem {loh} Y. L. Loh and E. W. Carlson, {\it Using inhomogeneity to raise the superconducting critical temperature in a two-dimensional XY model}, Phys. Rev. B {\bf 75}, 132506 (2007).

\bibitem{scalettar}
K. Aryanpour, E. R. Dagotto, M. Mayr, T. Paiva, W. E. Pickett and R. T. Scalettar, {\it Effect of inhomogeneity on $s$-wave superconductivity in the attractive Hubbard model}, Phys. Rev B {\bf 73}, 104518 (2006); K. Aryanpour, T. Paiva, W. E. Pickett and R. T. Scalettar, {\it $s$-wave superconductivity phase diagram in the inhomogeneous two-dimensional attractive Hubbard model}, Phys. Rev. B {\bf 76}, 184521 (2007).

\bibitem{Mishra2008} V. Mishra, P. J. Hirschfeld, and Yu. S. Barash, {\it Sublattice model of atomic scale pairing inhomogeneity in a superconductor}, Phys. Rev. B {\bf 78}, 134525 (2008).

\bibitem{andersen07} B. M. Andersen, P. J. Hirschfeld, A. P. Kampf, and M. Schmid, {\it Disorder-Induced Static Antiferromagnetism in Cuprate Superconductors}, Phys. Rev. Lett. {\bf 99}, 147002 (2007).

\bibitem{graser10} B. M. Andersen, S. Graser, and P. J. Hirschfeld,{\it Disorder-Induced Freezing of Dynamical Spin Fluctuations in Underdoped Cuprate Superconductors}, Phys. Rev. Lett. {\bf 105}, 147002 (2010).

\bibitem{mng13} M. N. Gastiasoro, P. J. Hirschfeld, and B. M. Andersen, {\it Impurity states and cooperative magnetic order in Fe-based superconductors}, Phys. Rev. B {\bf 88}, 220509(R) (2013).

\bibitem{astrid12} A. T. R\o mer, S. Graser, T. S. Nunner, P. J. Hirschfeld, and B. M. Andersen,
{\it Local modulations of the spin-fluctuation-mediated pairing interaction by impurities in d-wave superconductors}, Phys. Rev. B {\bf 86}, 054507 (2012).

\bibitem{foyevtsova2}K.  Foyevtsova, H. C. Kandpal, H. O. Jeschke, S. Graser, H.-P. Cheng, R. Valent\'{i} and P. J. Hirschfeld, {\it Modulation of pairing interaction in Bi$_2$Sr$_2$CaCu$_2$O$_{8+\delta}$ by an O dopant: A density functional theory study}, Phys. Rev. B {\bf 82}, 054514 (2010).

\bibitem{MillisSachdevVarma} A. J. Millis, S. Sachdev, and C. M. Varma, {\it Inelastic scattering and pair breaking in anisotropic and isotropic superconductors}, Phys. Rev. B {\bf 37}, 4975 (1988).

\bibitem{kemper09} A. F. Kemper, D. G. S. P. Doluweera, T. A. Maier, M. Jarrell, P. J. Hirschfeld, and H-P. Cheng
{\it Insensitivity of d-wave pairing to disorder in the high-temperature cuprate superconductors} Phys. Rev. B {\bf 79}, 104502 (2009).

\bibitem{Teknowijoyo} S. Teknowijoyo, K. Cho, M. A. Tanatar, J. Gonzales, A. E. B\"ohmer, O. Cavani,
V. Mishra, P. J. Hirschfeld, S. L. Bud'ko, P. C. Canfield, and R. Prozorov, {\it Enhancement of $T_c$ by point-like disorder and anisotropic gap in FeSe}, Phys. Rev. B {\bf 94}, 064521 (2016).

\bibitem{Scalapino95} D. J. Scalapino, {\it The case for $d_{x^2-y^2}$ pairing in the cuprate superconductors}, Physics Reports {\bf 250}, 329 (1995). 

\bibitem{roemerPMpairing} A. T. R\o mer, A. Kreisel, I. Eremin, M. A. Malakhov, T. A. Maier, P. J. Hirschfeld, and B. M. Andersen, {\it Pairing symmetry of the one-band Hubbard model in the paramagnetic weak-coupling limit: a numerical RPA study}, Phys. Rev. B {\bf 92}, 104505 (2015).

\bibitem{maier05} T.A. Maier, M. Jarrell, T. Pruschke, and M. Hettler, {\it Quantum cluster theories}, Rev. Mod. Phys. {\bf 77}, 1027 (2005).

\bibitem{hirschfeld86} P.J.  Hirschfeld, D. Vollhardt
and P. W\"olfle, Solid State Communications,  {\it Resonant
impurity scattering in heavy fermion superconductors} {\bf 59}, 111 (1986).


\bibitem{SchmittRink86} S. Schmitt-Rink, K. Miyake, and C. M. Varma, Phys. Rev.Lett. {\it Transport and Thermal Properties of Heavy-Fermion Superconductors:
A Unified Picture} {\bf 57}, 2575 (1986).


\bibitem{andersen06} B. M. Andersen, A. Melikyan, T. S. Nunner, and P. J. Hirschfeld, {\it Thermodynamic transitions in inhomogeneous d-wave superconductors}, Phys. Rev. B {\bf 74}, 060501(R) (2006).

\bibitem{suppl} See Supplementary Material



\bibitem{Andersen08} B. M. Andersen and P. J.
Hirschfeld, {\it Breakdown of Universal Transport in Correlated $d$-Wave Superconductors}, Phys. Rev. Lett. {\bf 100}, 257003 (2008).

\bibitem{Garg08} A. Garg, M. Randeria and N. Trivedi, {\it Strong correlations make high-temperature superconductors robust against disorder}, Nature Physics {\bf 4}, 762 (2008). 


\bibitem{tang16} S. Tang, V. Dobrosavljevi\'c, and E. Miranda, {\it Strong correlations generically protect d-wave superconductivity against disorder}, Phys. Rev. B {\bf 93}, 195109 (2016). 

\bibitem{chakraborty14} D. Chakraborty and A. Ghosal, {\it Fate of disorder-induced inhomogeneities in strongly correlated d-wave superconductors}, New Jour. Phys. {\bf 16}, 103018 (2014).

\bibitem{chakraborty17} D. Chakraborty, R. Sensarma, A. Ghosal {\it Effects of strong disorder in strongly correlated superconductors},	Phys. Rev. B {\bf 95}, 014516 (2017).

\bibitem{ghosal00} A. Ghosal, M. Randeria, and N. Trivedi,
{\it Spatial inhomogeneities in disordered d-wave superconductors}, Phys. Rev. B {\bf 63}, 020505 (R) (2000).
 
\bibitem{mng16} M. N. Gastiasoro, F. Bernardini, and B. M. Andersen, {\it Unconventional Disorder Effects in Correlated Superconductors} Phys. Rev. Lett. {\bf 117}, 257002 (2016).








\end{thebibliography}

\begin{thebibliography}{}
\bibitem{Andersen08} B. M. Andersen and P. J.
Hirschfeld, {\it Breakdown of Universal Transport in Correlated $d$-Wave Superconductors}, Phys. Rev. Lett. {\bf 100}, 257003 (2008).
\bibitem{Garg08} A. Garg, M. Randeria and N. Trivedi, {\it Strong correlations make high-temperature superconductors robust against disorder}, Nature Physics {\bf 4}, 762 (2008). 
\bibitem{astrid12} A. T. R\o mer, S. Graser, T. S. Nunner, P. J. Hirschfeld, and B. M. Andersen,
{\it Local modulations of the spin-fluctuation-mediated pairing interaction by impurities in d-wave superconductors}, Phys. Rev. B {\bf 86}, 054507 (2012).
\end{thebibliography}
\end{document}